\documentstyle[12pt]{article}
\newcommand{\beq}{\begin{equation}}
\newcommand{\eeq}{\end{equation}}
\newcommand{\bea}{\begin{eqnarray}}
\newcommand{\eea}{\end{eqnarray}}

\newcommand{\k}{\kappa}
\newcommand{\lb}{\label}
\newcommand{\e}{\epsilon}

\newcommand{\xp}{x}
\newcommand{\yp}{y}
\newcommand{\yip}{y_{i}}
\newcommand{\rg}{\sqrt{g}}

\newcommand{\tN}{\tilde{N}}
\renewcommand{\d}{\delta}
\renewcommand{\l}{{\cal M}}
\newcommand{\la}{\lambda}
\renewcommand{\L}{\Lambda}
\renewcommand{\b}{\beta}
\renewcommand{\a}{\alpha}
\newcommand{\E}{{\cal E}}

\newcommand{\ER}{\sqrt{\cal E}}

\newcommand{\A}{\mbox{\AE}}
\newcommand{\G}{{\cal G}}
\renewcommand{\H}{{\cal H}}
\newcommand{\N}{{\cal N}}

\newcommand{\R}{{\cal R}}
\newcommand{\ab}{\bar a}
\newcommand{\n}{\nu}
\newcommand{\m}{\mu}

\newcommand{\s}{\sigma}
\newcommand{\D}{\Delta}

\newcommand{\Th}{\Theta}

\newcommand{\oh}{\frac{1}{2}}

\newcommand{\non}{\nonumber}

\renewcommand{\t}{\tau}
\newcommand{\rf}[1]{(\ref{#1})}
\newcommand{\ra}{\rightarrow}

\newcommand{\pa}{\partial}

\begin{document}

\addtolength{\baselineskip}{0.20\baselineskip}

\hfill gr-qc/9610020


\hfill TIT/HEP-346/COSMO-79


\begin{center}

\vspace{24pt}

{ {\Large \bf The Mass Shell of the Universe} }

\end{center}

\vspace{12pt}
\begin{center}

\vspace{18pt}

{\sl A. Carlini}\footnote{E-mail: carlini@th.phys.titech.ac.jp}\\
{\small\em {Tokyo Institute of Technology, Oh-Okayama, Meguro-ku,
Tokyo 152, Japan}}\\[1.2em]

{\sl J. Greensite}\footnote{E-mail: greensite@theorm.lbl.gov~;
~greensit@stars.sfsu.edu}\\
{\small\em {Theoretical Physics Group, Ernest Orlando Lawrence Berkeley 
National Laboratory, \\
University of California, Berkeley, CA 94720 USA}}\\
and \\
{\small\em{Physics and Astronomy Dept., San Francisco State University, \\
San Francisco CA 94132, USA}}\\[1.2em]

\vspace{24pt}
\end{center}







\vspace{24pt}

\begin{center}

{\bf Abstract}

\end{center}

\bigskip

      The classical field equations of general relativity can be
expressed as a single geodesic equation, describing the free fall of a
point particle in superspace. Based on this formulation, a ``worldline''
quantization of gravity, analogous to the Feynman-Schwinger treatment of
particle propagation, is proposed, and a hidden mass-shell parameter is
identified.  We consider the effective action for the supermetric, which 
would be induced at one loop.  In certain minisuperspace
models, we find that this effective action is stationary for vanishing
cosmological constant.

\vfill

\newpage

\section{Introduction}

   In one of the classic papers of quantum electrodynamics, Feynman \cite{Feyn}
suggested that relativistic electron propagation could be understood in
terms of a sum over electron worldlines running both forwards and backwards
in time.  The evolution parameter was a path parameter,
associated with the proper
time of the electron worldlines, rather than the ``clock time'' of the
laboratory. Related ideas were discussed by Stueckelberg, Fock,
Nambu, and Schwinger \cite{Nambu}.  In this article we would like to
extend Feynman's worldline quantization of electron paths in
spacetime to the quantization of a closed Universe propagating in
superspace.

   The elements of the proper-time approach for relativistic particles
are, of course, very well-known.  Consider for simplicity a spinless
particle of mass $m$, propagating freely on a background spacetime
with metric $g_{\m \n}$.  The classical motion of the particle (i.e.
the geodesic equation) is derived from variation of the worldline
action
\bea
       S_p &=&  - m \int ds
\non \\
           &=& - m \int d\t \sqrt{-g_{\m \n} {dx^\m \over d\t}
                  {dx^\n \over d\t}}
\label{Sp}
\eea
Removing the square-root by introduction of a Lagrange multiplier
(lapse) $N$, we can write $S_p$ in the form
\beq
       S_p' = m \int d\t \; \left[ {1\over 2N} g_{\m \n} {dx^\m \over d\t}
                  {dx^\n \over d\t} - \oh N \right]
\label{Sp'}
\eeq
Applying the usual Legendre transform, one obtains the first-order form
\bea
      S_p'' &=& \int d\t \; \left[ p_\m {dx^\m \over d\t} - NH \right]
\non \\
     H  &=& {1\over 2m}(g^{\m \n}p_\m p_\n + m^2)
\label{Sp''}
\eea
Now go to the gauge $N=1$.  In this gauge, $\t=s$ is the proper-time
parameter of the classical equations of motion.  Adopting $s$ as the evolution
parameter for the quantized theory,
the amplitude for a relativistic
spinless particle to propagate from point $x'^\m$ to point $x^\m$ in an
interval $s$ can be expressed as a path-integral
\bea
      G(x,x';s) &=&  \int^x_{x'} Dx(s') \; \exp[{i\over \hbar}
                         \int_0^s  L_p' ds']
\non \\
       &=& <x|e^{-i\H_s s/\hbar}|x'>
\non \\
         L_p' &=& {m\over 2} \left( g_{\m \n} {dx^\m \over ds}
                  {dx^\n \over ds} - 1 \right)
\eea
Up to an operator-ordering, the corresponding Hamiltonian operator $\H_s$
describing state evolution in the evolution parameter $s$
\beq
      i\hbar {\pa \psi \over \pa s} = \H_s \psi
\label{Seq}
\eeq
is obtained from the classical Hamiltonian (with $N=1$) via the usual
replacement of c-number momenta by the corresponding operators, i.e.
\beq
      <x| \H_s |x'> = {1\over 2m}\left( -\hbar^2 \Box_x + m^2 -i\e +
            \xi R \right)  \d^D(x-x') [-g(x')]^{-1/2}
\label{H_S}
\eeq

   The Feynman propagator is proportional to the inverse of $\H_s$,
and the one-loop contribution to the gravitational effective action
\beq
         {\cal S}_{eff}[g_{\m \n}] = \oh i\hbar \mbox{Trln}[\H_s]
\label{Seff}
\eeq
is the trace logarithm of $\H_s$.  The proper-time formalism itself
has various uses, e.g. in calculating the Feynman propagator exactly
in certain, especially simple, background electromagnetic fields,
as well as in the evaluation of certain loop diagrams.  We note
that eigenvalues of the proper time Hamiltonian $\H_s$, such as those
used in the evaluation of the effective action, can take
on any value.  Classically, however, the mass-shell condition $H=0$
is to be respected (this follows from variation of \rf{Sp''} by $N$),
and for free particles this condition is imposed, in Dirac
quantization, as a constraint
on physical states $\H_s \psi = 0$.  For spinless particles, this
operator constraint is just the Klein-Gordon equation in curved spacetime.
In an interacting theory the mass-shell condition is relaxed somewhat;
it is required only of asymptotic states.
The 4-momentum of a virtual particle is allowed to violate the mass-shell
condition.

\section{The Worldline Action of a Closed Universe}

    We would now like to generalize the proper-time approach to the case
of gravity in combination with any number of interacting bosonic fields;
this calls for rewriting the gravitational action in the form
\beq
         S_g = - \l \int ds
\eeq
where $s$ is an invariant length parameter in the space of all fields
modulo spatial diffeomorphisms, i.e. superspace, and $\l$ is an arbitrary
dimensionless parameter. The only reasonable
candidate for $s$ is the usual action of general relativity, so the problem
is to reformulate that action as a proper time in superspace.  Such a
formulation was developed recently in ref. \cite{Me}; closely
related ideas were put forward long ago in ref. \cite{DeWitt}.   The
identification of action with proper time goes as follows:

   Let $\{q^a(x),p_a(x)\}$ represent a
set of gravitational and other bosonic fields, and their conjugate
momenta, with the fields scaled by an appropriate power of $\k^2=16\pi G_N$
so as to be dimensionless.\footnote{Only the metric formulation will be
considered here; hence the restriction to bosonic fields.} In a condensed
notation, the standard ADM action has the form
\bea
      S_{ADM} &=& \int d^4x \; [p_a \pa_0 q^a - N\H_x - N_i\H^i_x]
\non \\
      \H_x &=& \k^2 G^{ab} p_a p_b + \rg U(q)
\non \\
      \H^i_x &=& O^{ia}[q,\pa] p_a
\label{ADM}
\eea
where $G^{ab}$ and $U$ are, respectively, the metric and potential
in superspace, and the operator $O^{ia}$ is linear in the 3-d spacetime
covariant derivative.
Go to the ``shift gauge'' $N_i=0$.  The supermomentum constraints
$\H^i=0$ are not lost by this choice, since they are still required
for consistency of the Hamiltonian constraint with the equations of
motion.  Solving Hamilton's equation for the momenta in terms of
velocities, then solving the Hamiltonian constraint for the
lapse function $N$ in terms of velocities, and inserting both
expressions into $S_{ADM}$, one obtains the Baierlein-Sharp-Wheeler
(BSW) form of the gravitational action \cite{BSW}
\beq
      S_{BSW} = - \int d^4x \; \sqrt{-{1\over \k^2} \rg U G_{ab}
                            \pa_{0} q^a \pa_{0} q^b}
\label{BSW}
\eeq
in shift gauge $N_i=0$.  The BSW action is to serve as a proper-time
parameter.  It is also useful to introduce an arbitrary mass-scale $\s$
in order to define an evolution parameter $t$ with dimensions of time,
so that
\beq
        S_{BSW} = - \int ds = -\s \int dt
\label{t}
\eeq
Choose $x_0=t$. Comparing \rf{t} with \rf{BSW}, we have
\beq
   dt = {1\over \s} \int d^3x \; \sqrt{-{1\over \k^2} \rg U
                     G_{ab} dq^a dq^b }
\label{dt}
\eeq
Let $\tN$ denote the lapse function (derived by solving the Hamiltonian
constraint) associated with the time parameter $t$
\beq
      \tN = \sqrt{-{1\over 4\k^2 \rg U} G_{ab} \pa_t q^a \pa_t q^b }
\label{tN}
\eeq
Then we have
\beq
      1 = - {1\over \s} \int d^3x \; {1 \over 2 \tN \k^2}
                            G_{ab} \pa_t q^a \pa_t q^b
\label{1}
\eeq

   Now $t$ denotes a ``many-fingered'' time variable, with the different
possibilities distinguished by a choice of $\tN$.  Equation \rf{dt}
imposes only one global restriction on the choice of $\tN$.  From eq. \rf{tN}
we have
\beq
      \int d^3x \; \tN \rg U = \int d^3x \; \sqrt{-{1\over 4\k^2} \rg U
                     G_{ab} \pa_t q^a \pa_t q^b }
\eeq
which implies, from the definition \rf{dt}, the condition
\beq
      \int d^3x \; \tN \rg U = \oh \s
\label{conN}
\eeq
For a given $\tN$ satisfying this condition, there corresponds a time variable
proportional to $S_{BSW}$.  The condition is solved trivially by
\beq
     \tN = {\oh \s \N \over \int d^3x \; \N \rg U}
\label{calN}
\eeq
where $\N$ is unrestricted.  Inserting this form for $\tN$ back into
\rf{1}, we find
\beq
      1 = - {1\over \s^2} \int d^3x \; \left[ \int d^3x' \;
          \N \rg U \right] {1\over \N \k^2} G_{ab} \pa_t q^a \pa_t q^b
\eeq
or
\beq
     ds^2 = - \int d^3x \; \left[ \int d^3x' \;
          \N \rg U \right] {1\over \N \k^2} G_{ab} dq^a dq^b
\label{ds2}
\eeq

    Now introduce a mixed discrete/continuous ``coordinate index'' $(\a,x)$
in superspace:
\bea
     q^{(\a x)} \equiv q^{\a}(x) = \left\{ \begin{array}{cl}
                     \N(x)   & \a=0 \\
                     q^a(x)  & \a=a \ne 0 \end{array} \right.
\eea
Apart from notation we are extending the definition of superspace slightly
to include the non-dynamical field $\N(x)$, related via eq. \rf{calN}
to the lapse parameter.  Define a degenerate metric for
this extended superspace
\bea
  \G_{(a x)(b y)} &=&   \left[\int d^3x' \; \N \rg U \right]
          {1 \over \N(x) \k^2} G_{ab}(x) \d^3(x-y)
\non \\
  \G_{(0 x)(0 y)} &=& \G_{(a x)(0 y)} = \G_{(0 x)(b y)} = 0
\label{metric}
\eea
With these definitions, and an obvious summation convention,
eq. \rf{ds2} becomes
\beq
       ds^2 = - \G_{(\a x)(\b y)} dq^{(\a x)} dq^{(\b y)}
\eeq
The gravitational action then has the desired form
\bea
       S_g &=& - \l \int ds
\non \\
           &=& - \l \int d\t \; \sqrt{ - \G_{(\a x)(\b y)}
      {dq^{(\a x)}\over d\t} {dq^{(\b y)} \over d\t}  }
\label{Sg}
\eea

  Variation of the action $S_g$ w.r.t $q^{(\a x)}$ leads, in the usual
way, to a geodesic equation
\beq
     \G_{(\a x)(\b y)} {d^2 q^{(\b y)} \over d s^2} + \oh \left(
     {\d \G_{(\a x)(\b y)} \over \d q^{(\gamma z)} } +
     {\d \G_{(\a x)(\gamma z)} \over \d q^{(\b y)} } -
     {\d \G_{(\b y)(\gamma z)} \over \d q^{(\a x)} } \right)
     {dq^{(\b y)} \over ds}{dq^{(\gamma z)} \over ds}
     = 0
\label{geo}
\eeq
Identifying $ds=\s dt$, it is straightforward to verify that
the $\a \ne 0$ components of \rf{geo} are the equations
of motion
\beq
  {\pa \over \pa t}\left[ {1\over 2\tN \k^2} G_{ab} \pa_t q^b \right]
 - {1 \over 4\tN \k^2}{\pa G_{cd} \over \pa q^a}\pa_t q^c \pa_t q^d
+ \int d^3x' \; \tN {\d \over \d q^a(x)}(\rg U) = 0
\label{motion}
\eeq
while the $\a=0$ component is the Hamiltonian constraint
\beq
  {1 \over 4\tN^2 \k^2} G_{ab}\pa_t q^a \pa_t q^b + \rg U = 0
\label{constraint}
\eeq
These equations are identical to those obtained
from the ADM action \rf{ADM}, with the gauge choice $N_i=0$ and
$N=\tN$.  We have therefore interpreted the classical field equations
of general relativity as describing the free fall of a point particle
in superspace.\footnote{Eq. \rf{Sg} can also be motivated from Jacobi's
principle in mechanics, c.f. ref. \cite{Me}.}

   Some further comments are in order.  First, the choice of lapse
$N=\tN$ imposes only one global condition \rf{conN}
on the choice of lapse function. This does not result in any restriction
on the choice of foliation, but only on the time-label $t$ associated with each
hypersurface of a given foliation.  A second point is that
the degeneracy of the supermetric $\G_{(\a x)(\b y)}$ in
eq. \rf{metric} implies an infinite set of solutions for the geodesic
between any two points in superspace.  It is not hard to show that
these solutions are related by (ordinary D=4) time-reparametrizations, and
have the same ``proper time'' interval in superspace (proportional to $S_g$)
between those two points.  Finally, let us note that the parameter $\l$
in $S_g$, which is analogous to the mass parameter $m$ in the relativistic
particle action $S_p$, drops out of the classical configuration-space
equations of motion.

    Having recognized that the worldline action \rf{Sg} leads to the
same classical motion as the ADM action, we can proceed as in the
relativistic particle case to derive the proper-time Hamiltonian.
Again introducing a Lagrange multiplier $n$ to remove the square-root
\beq
    S_g' = \l \int d\t \; \left[ {1\over 2n}\G_{(ax)(by)}
{dq^{(ax)} \over d\t} {dq^{(by)} \over d\t} - \oh n \right]
\eeq
the first-order form is
\bea
     S_g''&=& \int d\t \; \left[ p_{(ax)} {dq^{(ax)} \over d\t} - nH \right]
\non \\
     H    &=& {1\over 2\l}[\G^{(ax)(by)} p_{(ax)} p_{(by)} + \l^2 ]
\non \\
          &=& {1\over 2\l}[\A + \l^2]
\label{1st_order}
\eea
where the supermetric
\beq
\G^{(ax)(by)}\dot ={\N \k^2 G^{ab} \delta^3(x-y) \over
\int d^3x' \; \N \rg U }
\eeq
and the expression
\bea
      \A &=&  \G^{(ax)(by)} p_{(ax)} p_{(by)}
\non \\
         &=&  { \int d^3x \; \N \k^2 G^{ab} p_a p_b \over
                \int d^3x' \; \N \rg U }
\eea
were introduced in ref. \cite{Us1}.  Variation of \rf{1st_order} with
respect to $q^a(x,\t),~p_a(x,\t)$ and $\N(x,\t),~n(\t)$ give us,
respectively, the set of Hamiltonian equations and constraints
\beq
      \pa_\t q^a(x) = {n\over 2\l}{\d \A \over \d p_a(x)} ~~,~~
      \pa_\t p_a(x) = -{n\over 2\l}{\d \A \over \d q^a(x)} ~~,~~
      {\d \A \over \d \N(x)} = 0 ~~,~~
          \A = -\l^2
\eeq
Setting $n=1$, so that $\t=s=\s t$, these equations are
equivalent to the usual Hamiltonian equations of
motion and Hamiltonian constraint
\bea
     \pa_t q^a(x) &=& \int d^3x' \; \tN {\d \over \d p_a(x)} \H
\non \\
     \pa_t p_a(x) &=& - \int d^3x' \tN {\d \over \d q^a(x)} \H
\non \\
      \H &=& {\k^2 \over \l} G^{ab} p_a p_b + \l \rg U = 0
\label{H2}
\eea
in the $N=\tN,~N_i=0$ gauge.  These equations of motion and (Hamiltonian)
constraint imply the remaining
supermomentum constraint as a consistency condition.

  The constant $\l$ is implicitly set to $\l=1$
in the usual Hamiltonian formulation of general relativity,
but we note at this point that there is no overwhelming reason
to make this choice.  The constant $\l$ appears as a
constant multiplicative factor in the worldline action \rf{Sg},
as does the mass $m$ in the worldline action \rf{Sp}.  Both of
these constants drop out of the corresponding geodesic equations.
Just as there is no way of determining the mass of a particle from
its trajectory in free fall, there is also no way of determining
the value of $\l$ from a given solution of the configuration-space
field equations.  In the context of the first-order formulation,
the condition $\A = -\l^2$ is in every sense analogous to the
particle mass-shell condition $g^{\m \n}p_\m p_\n = -m^2$. It
is therefore reasonable to identify $\l$ as a kind of (dimensionless)
mass-shell parameter, and to dignify the constraint $\A = -\l^2$
with the title ``mass-shell of the Universe''.

\section{Quantization}

    We now consider canonical quantization, in
the ``proper-time'' gauge $n=1$. The corresponding Schr\"odinger equation is
\bea
      i\hbar {\pa \Psi \over \pa s} &=& \H_s \Psi
\non \\
                  &=& {1\over 2\l} (\A + \l^2) \Psi
\label{Seq1}
\eea
which has the general $s$-dependent solution
\bea
       \Psi[q,s] &=& \sum_{\E \b} a_{\E \b} \Phi_{\E\b}[q]
              e^{i(\E-\l^2)s/(2\l \hbar)}
\non \\
       \A \Phi_{\E\b}[q] &=& -\E \Phi_{\E \b}[q]
\label{expand}
\eea
where the label $\b$ distinguishes among a linearly independent set
of eigenstates of $\A$ with eigenvalue $-\E$.
The classical constraint $\d \A /\d \N = 0$ becomes an operator
constraint ${\d \A \over \d \N} \Psi = 0$.  Inserting the eigenstate
expansion \rf{expand}, we find that each eigenstate $\Phi_\E$ satisfies
a Wheeler-DeWitt equation
\beq
      \left[ -{\hbar^2 \over \E} \k^2 ``G^{ab} {\d^2 \over \d q^a \d q^b}''
             + \rg U \right] \Phi_\E[q] = 0
\label{WD}
\eeq
associated with the parameter $\E$ (quotation marks indicate the
ordering ambiguity). Finally, if we also impose the mass-shell constraint
\beq
        \A \Psi = -\l^2 \Psi
\label{ms}
\eeq
then the only physical states are those with $\E = \l^2$, and the
(classically indeterminate) constant $\l$ can be absorbed, via
\beq
         \hbar_{eff} = {\hbar \over \l}
\eeq
into a rescaling of Planck's constant.

   There are two ways in which the off mass-shell states, with
$\E \ne \l^2$, may be physically relevant.  First, the
mass-shell constraint \rf{ms} may not really {\it be} a constraint,
at the quantum level.   The mass-shell
condition is derived  by trading the square-root form of the action for an
expression involving a Lagrange multiplier.  What if one avoids this
step, and quantizes the square-root action $S_{BSW}$ directly?  This approach
has been advocated in ref. \cite{Us1,Us2}, and it leads to a formulation
in which the dynamical equation \rf{Seq1} is supplemented by the
constraints $(\d \A / \d \N )\Psi = 0$, but without the mass-shell
constraint $\A = -\l^2$.  It should be noted, once again, that there is no way
to determine $\l$ classically, or to verify the mass-shell
condition $\A=-\l^2$, since the configuration-space equations are
independent of $\l^2$.  Determination of $\l^2$ would require
a violation of the Einstein field equations; it is analogous to
trying to determine the mass of a particle from its trajectory
in free fall.  Moreover, the freedom to choose arbitrary foliations of
4-space is already reflected in the constraint $(\d \A / \d \N) \Psi = 0$.
In the formulation of \cite{Us1,Us2}, the physical Hilbert space is spanned
by the solutions of a family of Wheeler-DeWitt equations \rf{WD}, one
equation for each eigenvalue $-\E$ of $\A$.

   The second way in which off mass-shell states could become relevant
is suggested by the phenomenon of black hole evaporation.
Although it is known that black holes must lose mass via Hawking radiation,
it is not known what the final state of the radiative process
will be.  It is possible that the black hole disappears entirely, and
this might be considered a case of topology change involving the production of
a ``baby universe'', analogous to similar processes in string theory.
It is also possible that the evaporation is not complete, and the
black hole leaves a remnant.  Let us suppose that the first alternative,
namely, complete evaporation accompanied by baby universe production, is
the correct one.  In that case the Universe is {\it not} really in
free fall; there will be interactions associated with topology changing
processes (emission and absorbtion of baby universes).

    A satisfactory description of topology-changing processes awaits
development of a ``third-quantized'' theory of gravity \cite{3q}; unfortunately,
at present, we do not even have a satisfactory understanding of
second-quantized gravity.  Still, it may be possible to obtain some
insight into ``multi-versal'' effects via the worldline formulation.  For
example, by direct analogy to eq. \rf{Seff}, the 1-loop contribution of
virtual universe loops to the effective action would be
\beq
      S_{eff}[\G_{ab}] = {i\hbar \over 2} \mbox{Trln}[\A + \l^2]
\eeq
where the trace runs over a basis of states $\Phi_\E$ satisfying the
one-parameter family of Wheeler-DeWitt equations \rf{WD}.  Of course,
the supermetric $\G_{ab}$, unlike the ordinary spacetime metric
$g_{\m \n}$, is not arbitrary; it is constrained to have the form
\rf{metric}.  Therefore $S_{eff}$ may be regarded as a functional of
the potential term $U(q)$.  But the form of $U(q)$ is also tightly
constrained: it is the sum of all possible potential terms that could
appear in an ADM Hamiltonian.  With this restriction, $S_{eff}$ is
just a function of the coupling constants of each possible interaction
term, i.e.
\beq
        S_{eff}[\G_{ab}] = S_{eff}[\la,e^2,g^2,...]
\eeq
and the couplings are now viewed as dynamical variables.
Variation of $S_{eff}$ with respect to the couplings could,
in principle, determine their phenomenological
values, very much in the spirit of Coleman's
``Big Fix'' \cite{Coleman}.

   Let us illustrate this possibility with a minisuperspace toy model,
in which the supermetric $\G_{ab}$ depends on one parameter only, namely,
the cosmological constant $\la$.  The starting point is the minisuperspace
action representing a closed, homogenous and isotropic
Friedman-Robertson-Walker (FRW) universe filled with a three-component,
minimally coupled scalar field $\vec{\phi}\dot =(\phi_1, \phi_2, \phi_3)$, i.e.
\beq
     S = \oh \int dt \left[ -{a\dot{a}^2\over N}
         + {a^3\dot{\vec{\phi}} \cdot \dot{\vec{\phi}}\over N}
          + N(a - \la a^3) \right]
\label{adm1}
\eeq
where the 4-d invariant length is
\beq
ds^2={\hat\sigma}^2(-N^2dt^2+a^2d\Omega_3^2)
\eeq
and with ${\hat\sigma}^2\dot = 2G_N/3\pi$.

  With the choice of coordinates $q^0=a,~q^i=\phi^i$, the
supermetric for the corresponding worldline action
\beq
     S_g = - \l \int d\t \sqrt{-\G_{ab} \dot{q}^a \dot{q}^b}
\label{miniwld}
\eeq
reads
\beq
\G_{ii}=-a^2\G_{00} =a^4(\la a^2 -1)~~~~;~~~~i=1, 2, 3
\label{metric1}
\eeq
Now, on general grounds of diffeomorphism invariance in minisuperspace,
the effective action for a generic FRW universe will have a weak-curvature
(adiabatic) expansion of the form
\beq
      S_{eff}[\G_{ab}] = \int da\; \int d\vec{\phi}~\sqrt{|\G|} \left[
            \L_S + \k_S {\cal R} + O({\cal R}^2) \right]
\label{adiabat1}
\eeq
where $\L_S$ and $\k_S$ are the
(dimensionless) ``supercosmological constant''
and ``super Newton's constant'', respectively, and ${\cal R}$ is the scalar
``supercurvature'' of $\G_{ab}$.
In general, since $\L_S,~\k_S$ are divergent at one loop,
even in simple minisuperspace models, it must be assumed that either
these constants are renormalized (and there exists a bare
action $S_0[\G_{ab}]$), or else that there is a fundamental cutoff of some
kind in the theory.

Let us now temporarily compactify the ranges of integration in
\rf{adiabat1} so that the scale factor runs from $a=0$ to
$a=\bar a$ and the scalar fields run from $\phi_i=-\phi_{i M}$ to
$\phi_i=\phi_{i M}$, and keep only the leading term in the adiabatic expansion
\rf{adiabat1}.
Then the effective action \rf{adiabat1} reads
\bea
S_{eff}&\simeq &\L_S \int_0^{\ab} da\int_{-\phi_{1M}}^{\phi_{1M}} d\phi_1
\int_{-\phi_{2M}}^{\phi_{2M}} d\phi_2 \int_{-\phi_{3M}}^{\phi_{3M}} d\phi_3
~a^7(\la a^2-1)^2
\non \\
&=&\left( \int d^3\phi \right)\L_S\ab^8\left (
{\lambda^2 \ab^4\over 12}-{\lambda \ab^2\over 5}+{1\over 8}\right)
\lb{r22b}
\eea
It is easy to check that $S_{eff}$ is stationary at
\beq
{d S_{eff} \over d\la}= 0 ~~~ \Longrightarrow ~~
\la\simeq {6\over 5\ab^2}
\lb{r29b}
\eeq
with the result that $\la \ra 0^+$ as $\ab\ra \infty$.
It is also straightforward to show that this stationary point is actually a
minimum for $S_{eff}$ provided that $\L_{S}>0$.

\section{Inclusion of Mass Terms and Supercurvature}

     Any minisuperspace model is a toy, and
only illustrates effects which might (or might not) be present
in the full theory.  Still, even within the category of toy models, it is
interesting to study whether the vanishing of the cosmological
constant survives some modest complications of the minisuperspace
action, and/or improvements in the approximations for \rf{adiabat1},
e.g., the inclusion of contributions from the supercurvature terms in the
adiabatic expansion of the effective action.

   We consider the action for a FRW universe
filled with $N_s$ scalar fields $(\phi_1, .., \phi_{N_s})\dot = \vec{\phi}$
with potential $V(\phi)$, i.e.
\beq
S={1\over 2}\int dt\left \{ -{a {\dot a}^2\over N}+{a^3{\dot{\vec{\phi}}}\cdot
{\dot{\vec{\phi}}}\over N}+Na[1-(\la +V(\phi))a^2]\right \}
\lb{f9a}
\eeq
Again choosing coordinates $q^0=a, ~q^i=\phi^i$, then, from eq. \rf{miniwld},
the diagonal, $N_s+1$-dimensional worldline supermetric $\G_{ab}$ is just
\beq
\G_{ii}=-a^2\G_{00}=a^4[(\la +V(\phi))a^2-1]~~~;~~~i=1, ... N_s
\lb{f9b}
\eeq
and the effective action is given by eq. \rf{adiabat1}.
The question is whether the stationary point of the effective action
\rf{adiabat1}, with supermetric \rf{f9b}, is still at $\la = 0^+$.
We will now consider some cases for various numbers of scalar fields, with
and without mass-term potentials.

\subsection{$N_s$ massless scalar fields}

As a first example, we consider the model of a FRW universe filled with
$N_s$ massless, minimally coupled scalar fields, i.e. the case with potential
\beq
V(\phi)=0
\lb{f0}
\eeq
For this scalar potential, inserting the supermetric \rf{f9b} into eq.
\rf{adiabat1} we can easily write down for the effective action
(neglecting the `supercurvature' contributions)
\bea
S_{eff}&\simeq &\L_S\int_0^{\ab} da\int_{-\phi_{1M}}^{\phi_{1M}} d\phi_1 ~...~
\int_{-\phi_{N_sM}}^{\phi_{N_sM}} d\phi_{N_s}~a^{2N_s+1}|\la a^2-1|^{(N_s+1)/2}
\non \\
&=&{N_s!\L_{S} \left (\prod_{i=1}^{N_s}I_{\phi_i}\right )\ab^{2(N_s+1)}
\over 3(3N_s+1)!!~\xp^{N_s+1}}\biggl \{ 2^{N_s}(N_s-1)!! + [\Th(\xp-1)
\non \\
&-&\Th(1-\xp)]\xp^{N_s} |\xp-1|^{(N_s+3)/2}
\sum_{k=0}^{N_s} 2^k{(3N_s-2k+1)!!\over (N_s-k)!}\xp^{-k}\biggr \}
\lb{f10}
\eea
where $\Th(x)$ is the Heaviside step function and the quantities
$\xp$ and $I_{\phi_i}$ are given by
\beq
\xp\dot =\la\ab^2
\label{g5}
\eeq
and
\beq
I_{\phi_i}\dot =\int_{-\phi_{iM}}^{\phi_{iM}}d\phi_{i}
\lb{f10a}
\eeq


Taking the derivative of the effective action \rf{f10} with respect to
$\la$ we get
\bea
{dS_{eff}\over d\la}&=&
{2^{N_s}(N_s+1)!\L_{S}\left(\prod_{i=1}^{N_s} I_{\phi_i}\right )
\ab^{2(N_s+2)}\over 3(3N_s+1)!!~\xp^{N_s+2}}\biggl \{-(N_s-1)!!
\non \\
&+&|\xp-1|^{(N_s+1)/2}\sum_{k=0}^{N_s+1} 2^{-k}{(N_s+2k-1)!!\over k!}
\xp^{k}\biggr \}
\lb{f11}
\eea
Unfortunately, the stationarity condition coming from eq. \rf{f11}, i.e. by
imposing $dS_{eff}/d\la=0$, cannot be easily solved for arbitrary $N_s$.
However, one can still prove the existence of a finite number
(at least one) of stationary points of $S_{eff}$ which are all at $\xp >0$
and at a finite distance from the origin $\xp=0$.
In fact, studying the behaviour of $dS_{eff}/d\la$ in the range
$\xp\geq 1$, we get (for $\L_S>0$)
\bea
{dS_{eff}\over d\la}(\xp=1)&=&-{2^{N_s}(N_s+1)!!\L_{S}\left(\prod_{i=1}^{N_s}
I_{\phi_i}\right )\ab^{2(N_s+2)}\over 3(3N_s+1)!!}~<0
\non \\
{dS_{eff}\over d\la}(\xp\rightarrow +\infty)&\sim &
{\L_{S}\left(\prod_{i=1}^{N_s} I_{\phi_i}\right )
\ab^{2(N_s+2)}\xp^{(N_s-1)/2}\over 6}~>0
\lb{f15}
\eea
(with the inequality signs reversed in the case $\L_S<0$).
On the other hand, it is possible to check (i.e., by using Mathematica), that
\beq
{dS_{eff}\over d\la}~<0 ~~~;~~~ \forall ~\xp \leq 0
\lb{f21}
\eeq
(again with the inequality sign reversed in the case $\L_S<0$).
In other words, eqs. \rf{f15} and \rf{f21} imply that $dS_{eff}/d\la$
will have at least one finite zero at $\xp\dot =x_{1}>1$, and at most a
finite number of extra zeros at $\xp\dot =x_n=finite >0$.
Therefore, the effective action $S_{eff}$ will be stationary at
\beq
{d S_{eff} \over d\la}\biggl\vert_{x_{1}} = 0 ~~~ \Longrightarrow ~~
\la={x_{1}\over \ab^2}
\lb{f14}
\eeq
(or, at any other of the points $x_{n}=c_n x_{1}$,
with $c_n=constant>0$).
Removing the cutoff, $\ab\rightarrow \infty$, this leads again to the result
that $\la=0^+$.

\paragraph{$N_s=0$ scalar fields}

  In this case the minisuperspace is one-dimensional, with the single
metric component
\beq
\G_{00}=-a^2(\la a^2-1)
\lb{metric2}
\eeq
The supercurvature $\R$ is, of course, identically zero.
One can then immediately write the effective action from eq. \rf{adiabat1} as
\bea
    S_{eff}[\lambda]
&=& \L_S \int_0^{\ab} da \; a |\la a^2 -1|^{1/2}
\non \\
&=& {\L_{S}\ab^2\over 3\xp}\{1+[\Th(\xp-1)-\Th(1-\xp)]|\xp-1|^{3/2}\}
\label{g4}
\eea
Taking the derivative of \rf{g4} with respect to $\la$, one obtains
\beq
{d S_{eff} \over d\la} = 0 ~~~ \Longrightarrow ~~
\la={[(3+2\sqrt{2})^{1/3}+(3-2\sqrt{2})^{1/3}-1]\over \ab^2}
\label{g8}
\eeq
with the result that $\la\rightarrow 0^+$ as the regulator $\ab$ is removed.
It is also straightforward to check that this stationary point is a minimum for
the effective action if $\L_S>0$.

\paragraph{$N_s=1$ massless scalar field (with supercurvature contribution)}

Next we consider the case of a single massless scalar field.
In this case the supermetric will have two independent diagonal entries,
$\G_{00}$ and $\G_{11}$, which can again be read from eq. \rf{f9b},
and we can also improve the evaluation of the effective action by including
the contribution of the supercurvature term given by
\beq
\R=-{4\la\over a^2(\la a^2-1)^3}
\lb{r2}
\eeq
The effective action, up to first order contributions from the
adiabatic expansion in $\R$, reads
\bea
S_{eff}&\simeq &\int_0^{\ab} da\int_{-\phi_{M}}^{\phi_{M}} d\phi \left [
\L_S~a^3|\la a^2-1|-4\k_S~{\la a |\la a^2-1|\over (\la a^2-1)^3}\right ]
\non \\
&=&I_{\phi}\biggl \{\L_{S}{\ab^4\over 6}\biggl[\left (\xp-
{3\over 2}+{1\over \xp^2}\right)\Th(\xp-1)
+\left ({3\over 2}-\xp\right)\Th(1-\xp)\biggr ]
\non \\
&-&{\k_{S}\over (\xp-1)}[(\xp-3)\Th(\xp-1)
+2\xp\Th(1-\xp)]\biggr\}
\lb{r3}
\eea
Evaluating the derivative of \rf{r3} with respect to $\la$ we get
\bea
{dS_{eff}\over d\la}
&\simeq &{I_{\phi }\L_{S}\ab^6\over 6}
\biggl \{\left [{(\xp^3-2)\over \xp^3}-{\a\over (\xp-1)^2}\right]\Th(\xp-1)
\non \\
&-&\left [ 1-{\a\over (\xp-1)^2}\right ]\Th(1-\xp)\biggr\}
\lb{r8}
\eea
where we have introduced the quantity
\beq
\a\dot ={12\k_{S}\over \L_{S}\ab^4}
\lb{r11}
\eeq
Now, provided that $\L_{S}\not =0$,\footnote{When $\L_S=0$
the analysis of the stationary points of $S_{eff}$ critically depends
on the relative scaling between $\la$ and the cutoff $\ab$, and
is not conclusive.} imposing the stationarity
condition with $dS_{eff}/d\la$ given by eq. \rf{r8}, and noting from
eq. \rf{r11} that the contribution coming from the supercurvature term can be
neglected in the limit when the regulator for the scale factor is removed
($\ab\rightarrow \infty$), it is straightforward to get the result
\beq
{d S_{eff} \over d\la}= 0 ~~~ \Longrightarrow ~~
\la\simeq {2^{1/3}\over \ab^2}
\lb{r15}
\eeq
In other words, the effective action is stationary at $\la=0^+$
as the regulator $\ab\rightarrow \infty$ is removed.
Moreover, evaluating the second order derivative of $S_{eff}$ with
respect to $\la$ it is easy to
see that the stationary point is a minimum for $S_{eff}$ if $\L_S>0$.

\paragraph{$N_s=3$ massless scalar fields (with supercurvature contribution)}

The analysis of the model of a FRW universe filled with three massless
scalar fields essentially proceeds along the same lines
as in the previous paragraph.
In particular, the supermetric now has two extra diagonal elements
(i.e., $\G_{22}=\G_{33}\equiv\G_{11}$,
plus $\G_{00}$, all readable from eq. \rf{f9b}), and  the supercurvature
turns out as
\beq
\R={6[9(\la a^2)^2-14\la a^2+4]\over a^4(\la a^2-1)^3}
\lb{r21}
\eeq
so that the effective action, up to first order contributions from
$\R$, reads
\bea
S_{eff}&\simeq &\int_0^{\ab} da\int_{-\phi_{1M}}^{\phi_{1M}} d\phi_1
\int_{-\phi_{2M}}^{\phi_{2M}} d\phi_2 \int_{-\phi_{3M}}^{\phi_{3M}} d\phi_3
\biggl \{\L_S~a^7(\la a^2-1)^2
\non \\
&+&6\k_S~{a^3 [9(\la a^2)^2-14\la a^2+4]
\over (\la a^2-1)}\biggr \}
\non \\
&=&\left(\prod_{i=1}^3 I_{\phi_i}\right)\ab^4\biggl [\L_{S}\ab^4\left (
{\xp^2\over 12}-{\xp\over 5}+{1\over 8}\right)
\non \\
&+&3{\k_{S}\over \xp^2}
\left (3\xp^3-{5\over 2}\xp^2-\xp-\ln|\xp-1| \right )\biggr ]
\lb{r22}
\eea
Finally,
taking the derivative with respect to $\la$ we get
\bea
{dS_{eff}\over d\la}
&\simeq& \left(\prod_{i=1}^3I_{\phi_i}\right )\L_{S}\ab^{10}
\biggl \{{\xp\over 6}-{1\over 5}
\non \\
&+&{3\over 4}\a \left [1+{(\xp-2)\over 3\xp^2(\xp
-1)}+{2\over 3\xp^3}\ln|\xp-1|\right]\biggr \}
\lb{r24}
\eea
where $\a$ is again defined according to eq. \rf{r11}.
Let us consider the case $\L_{S} \not =0$ first.
In this ansatz, using similar arguments to those of the previous section one
can easily see that the contribution coming from the supercurvature term
(the $\a$ term in eq. \rf{r24}) can be neglected when removing the cutoff
$\ab$,
and therefore the stationarity condition for $S_{eff}$ implied by eq.
\rf{r24} becomes
\beq
{d S_{eff} \over d\la}= 0 ~~~ \Longrightarrow ~~
\la\simeq {6\over 5\ab^2}
\lb{r29}
\eeq
Eq. \rf{r29} again predicts the value $\la =0^+$ as $\ab \rightarrow \infty$.
Contrarily to the previous model for a single scalar field, in the three
massless scalar field case we can also consider the ansatz $\L_{S}=0$.
In this case, in fact, it is easy to check from eq. \rf{r24} that the
stationarity condition for $S_{eff}$ has a solution at the point $x\dot =x_{2}=
finite >0$
\beq
{d S_{eff} \over d\la}\biggl\vert_{x_{2}} = 0 ~~~ \Longrightarrow ~~
\la\simeq {x_{2}\over\ab^2}
\lb{r31}
\eeq
In other words, also in this case $\la=0^+$ is a stationary
point for the effective action.
Finally, evaluating
the second order derivative of $S_{eff}$ with respect to $\la$,
at the stationary points \rf{r29} or \rf{r31}, it is
easy to check that these are minima for $S_{eff}$ either if $\L_{S}>0$
(for any $\k_{S}$) or if $\k_{S}>0$ (when $\L_{S}=0$).

\subsection{$N_s=4r-1$ massive scalar fields}

The next complication of the FRW universe toy model is to consider the case
of an odd number $N_s=4r-1$ ($r=1, 2, ..$) of massive, minimally coupled
scalar fields (the case of one single massive scalar field
is separately treated in the Appendix) with potential
\beq
V(\phi)=\sum_{i=1}^{4r-1}m_i^2\phi_i^2
\lb{f1}
\eeq
The supermetric is once again diagonal, with $4r-1$ identical entries
$\G_{ii}$ plus $\G_{00}$ (which can be easily read off eq.
\rf{f9b}), and there is no ambiguity in the sign of its determinant when
evaluating $S_{eff}$.
In particular, making use of the binomial expansion theorem three times,
the effective action can be written (neglecting the `supercurvature'
contributions) as
\bea
S_{eff}&\simeq &\L_S\int_0^{\ab} da\int_{-\phi_{1M}}^{\phi_{1M}} d\phi_1 ~..
\int_{-\phi_{N_sM}}^{\phi_{N_sM}} d\phi_{N_s}
~a^{2N_s+1}\left [\left(\sum_{i=1}^{N_s}m_i^2\phi_i^2 +\la\right )a^2-1\right
]^{2r}
\non \\
&=&{1\over 2}({N_s+1\over 2})!\L_{S}
\left (\prod_{i=1}^{N_s}I_{\phi_i} \right )\ab^{2(N_s+1)}
\non \\
&\times&\sum_{k=0}^{2r} \sum_{j=0}^k \sum_{s_1 .. s_{N_s}=0}^{k-j}
{\delta\left(\sum_{i=1}^{N_s}s_i-k+j\right )(-1)^k\xp^j y_{1}^{s_1}...
y_{N_s}^{s_{N_s}}\over (2s_1+1)s_1!...(2s_{N_s}+1)s_{N_s}!(2r-k)!j!(4r+k)}
\lb{f3}
\eea
where we have used the `cosmological constant-variable' defined by eq.
\rf{g5} and introduced the new `mass-variable' $\yip$ according to
\beq
\yip\dot =m_{i}^2\phi_{i M}^2\ab^2
\lb{f5}
\eeq

Now, in order to find the stationary points of $S_{eff}$, we can simplify
the whole analysis by taking partial derivatives with respect to $\la$
and $m_{i}^2$ and evaluating them at $\yip=0$ (i.e., at zero masses for
the scalar fields $\phi_i$).
Proceeding in this way and noting that the only relevant terms
surviving at $\yip=0$ from the sums in eq. \rf{f3} are, for the
derivative with respect to $m_{i}^2$, those with $j=k-1$, and,
for the derivative with respect to $\la$ and the effective action itself,
those with $j=k$, we obtain the formulas
\bea
{\partial S_{eff}\over \partial (m_{i}^2)}\biggl\vert_{\yip=0}&=
&{\phi_{i M}^2\over 3}{\partial S_{eff}\over \partial \la}\biggl
\vert_{\yip=0}={\phi_{i M}^2\over 3}
{\partial \over \partial \la} \left [S_{eff}
\vert_{\yip=0}\right ]
\non \\
&\simeq &{\L_{S}
\ab^{2(4r+1)}(I_{\phi_i})^2\left (\prod_{j=1}^{4r-1}I_{\phi_j}\right )
\over 24}
\non \\
&\times&
\sum_{k=0}^{2r}(-1)^k \left (\begin{array}{c} 2r \\ k \end{array}
\right){k\over (4r+k)}\xp^{k-1}
\lb{f23}
\eea

The effective action \rf{f3} evaluated at $\yip=0$ is, of course,
the same as that considered in section 4.1 for the case of $N_s$ massless
scalar fields (with the restriction that $N_s=4r-1$), and as a consequence
also the stationarity conditions derived from eqs. \rf{f23} are equivalent
to the massless model condition coming from eq. \rf{f11}.
Then the result is that also for the massive model considered here there
is at least one (trivial) stationary point at $m_{i}=0$ ($i=1, 2, .. 4r-1$)
and $\la$ given by eq. \rf{f14}.


Moreover, since the general stationarity conditions which one would derive by
equating the partial derivatives of $S_{eff}$, eq. \rf{f3}, with respect to
$\la$ and $m_{i}^2$ for {\it arbitrary} $m_{i}$ are still polynomial
equations of finite order in $\xp$ and $\yip$, it is easy to see that any
other eventual stationary point for the effective action would still be at
$|x_{n}|=finite~~,~~ |y_{i, n}|=finite$.
Therefore, we can again conclude that the stationary point for the effective
action representing a FRW universe filled with $N_s=4r-1$ massive scalar fields
is, after removal of the cutoffs, unique, i.e. at $|\la|=0$ and $m_{i}=
0$ ($i=1, 2,  .. 4r-1$).\footnote{The modulus in the value for $\la$
is actually due to our ignorance about the signs of the other eventual
stationary points $x_n$ and $y_{i, n}$.}

\paragraph{$N_s=3$ massive scalar fields}

In the case $N_s=3$, the algebra is especially simple.
In this case the effective action \rf{f3} simplifies to
\bea
S_{eff}&=&\L_{S}\ab^8\left (\prod_{i=1}^{3}I_{\phi_i} \right )
\biggl [ {1\over 12}
\biggl ({1\over 5}\sum_{i=1}^3\yip^2+{2\over 9}\sum_{i>j=1}^3\yip y_{j}
\non \\
&+&{2\over 3}\xp\sum_{i=1}^3\yip+\xp^2\biggr )-{1\over 5}\left ({1\over 3}
\sum_{i=1}^3\yip+\xp\right )+{1\over 8}\biggr ]
\lb{r48}
\eea
In particular, the partial derivatives with respect to $\la$ and $m_{i}^2$
turn out as
\beq
{\partial S_{eff}\over \partial (m_{i}^2)}=
{\L_{S}\ab^{10}\left (\prod_{j=1}^{3}I_{\phi_j}
\right )(I_{\phi_i})^2\over 120}\left [\yip+
{5\over 9}\sum_{j\not =i}y_j +5{\xp\over 3}-2\right ]
\lb{r49}
\eeq
\beq
{\partial S_{eff}\over \partial \la}=
{\L_{S}\ab^{10}\left (\prod_{i=1}^{3}I_{\phi_i}
\right )\over 6}\left [{1\over 3}\sum_{i=1}^3\yip
+\xp -{6\over 5}\right ]
\lb{r50}
\eeq
from which it is straightforward to find that the unique stationary point
of $S_{eff}$ is at
\bea
{\partial S_{eff} \over \partial\la}= 0 ~~~ \Longrightarrow ~~
\la&=&{6\over 5\ab^2}
\non \\
{\partial S_{eff} \over \partial(m_{i}^2)}= 0 ~~~ \Longrightarrow ~~
m_{i}^2&=&0~~~;~~~i=1, 2, 3
\lb{r51}
\eea
On removing the cutoffs we find, as anticipated in the last section,
that the stationary point is at $\la=0^+$ and $m_{i}=0$ ($i=1, 2, 3$).
We can also check the nature of this stationary point by evaluating the
eigenvalues of the Hessian of $S_{eff}$, finding
that the stationary point \rf{r51} is a minimum for $S_{eff}$ provided
$\L_{S}>0$.

\section{A New Source of Decoherence?}

   We have speculated that the dynamics of the Universe is
{\it not} precisely free fall, possibly due to topology-changing
absorbtion/emission processes.  If so, then in the interval
between such interactions the Universe propagates as a virtual particle in
superspace.  Alternatively, as we have suggested in some previous articles,
the mass-shell constraint may not really {\it be} a constraint at
the quantum level. In either case, the Universe would be propagating
somewhat off-shell.  It is interesting to imagine how this off-shell character
might manifest itself, if the effect would be large enough to be
observable.

   Consider  a solution of the evolution equation \rf{Seq1} and constraints
\hfil\break
$(\d \A / \d \N )\Psi = 0$, which is a superposition of two WKB states
\beq
      \Psi(q,\t) = \Psi_A(q,\t) + \Psi_B(q,\t)
\eeq
of the form
\bea
      \Psi_A(q,\t) &=& \int d\E D\a \; F_A(\E,\a) \exp\left[ {i\over \hbar}
 \left\{ (\E - \l^2)\t - \ER S[Q,\a] \right\} \right] \phi_A(q)
\non \\
      \Psi_B(q,\t) &=& \int d\E D\a \; F_B(\E,\a) \exp\left[ {i\over \hbar}
 \left\{ (\E - \l^2)\t - \ER S[Q,\a] \right\} \right] \phi_B(q)
\non \\
\label{AB}
\eea
where $\t = s/2\l$ is the rescaled evolution (proper-time) parameter
and $F_{A,B}$ are
distributions concentrated at $\E=\l^2$ (with a rms uncertainty
$\D \E$) and at parameter values $\{\a\}=\{\a_{A,B}\}$ respectively.
The functional $S[Q,\a]$ is a solution, invariant under
3-space diffeomorphisms, of the Hamilton-Jacobi equation
\beq
       \k^2 G^{ij} {\d S \over \d Q^i(x)}{\d S \over \d Q^j(x)}
           + \rg {\cal U}[Q(x)] = 0
\eeq
with $\{ \a \}$ a set of integration constants.
In these equations $Q$ represents the set of degrees of freedom to be
treated semiclassically, and $\rg {\cal U}[Q]$ is the part of the
superpotential involving only those degrees of freedom.  Note that in the case
of on-shell propagation, i.e. $\E=\l^2$, the $\t$-dependence drops out
of the wavefunction, and the expressions in \rf{AB} are just WKB solutions
of the Wheeler-DeWitt equation.

   Let us imagine that in some region of superspace where the amplitudes
$\Psi_{A,B}$ are non-negligible, the phase difference
\beq
     \d S[Q'] = \left| S[Q,\a_A] - S[Q,\a_B] \right|
\eeq
depends mainly on a small subset $Q'$ of the $Q$ degrees of freedom.
For example, $Q'$ might refer to the location of a particle recorded
on a photographic plate, and $\d S$ refers to the difference in
action, associated with two well separated particle paths in an
interferometer, leading to the same final location.

   We now ask whether the two components $\Psi_A$ and $\Psi_B$ will interfere
coherently, in the sense that the term is used in optics, in a measurement
of $Q' \subset Q$.  If $\D \E \ne 0$, then we must consider
stationarity with respect to variation
in $\E$, as well as stationarity with respect to variations in the
parameters $\a$.  The stationary phase condition tells us
that the components $\Psi_{A}$ and $\Psi_{B}$ are peaked at a
given configuration $Q$ at parameter times
\beq
      \t_A = {S[Q,\a_A] \over 2 \ER } ~~~~, ~~~~
      \t_B = {S[Q,\a_B] \over 2 \ER }
\eeq
respectively, with $\E$ evaluated at $\E=\l^2$.
Interference of wavefunctions $\Psi_A$ and $\Psi_B$
is coherent, in the sense of physical optics, if the relative phase
between the two wavefunctions is constant in the $\t$-interval
$[\t_A,\t_B]$.  In standard terminology the ``linewidth'' of
the wavefunction is $\D \E /\hbar$, and the ``coherence time'' is
$\D \t = \hbar / \D \E$.  If the linewidth has a stochastic origin, then
the phase of the wavefunction at $\t + \D \t$  is not related in a simple
way to the phase at parameter time $\t$.
The coherence criterion is then
\beq
        \d \t < \D \t = {\hbar \over \D \E}
\eeq
where
\beq
       \d \t \equiv |\t_A - \t_B| \approx {1\over 2\ER} \d S[Q']
\eeq
which means
\beq
      {1\over 2\ER} \d S < {\hbar \over \D \E}
\eeq
Defining $\hbar_{eff}(\E) = \hbar /\ER$ and the dispersion
\beq
     \d \hbar = \left| {d \over d \E} \hbar_{eff} \right|_{\E=\l^2} \D \E
              = \oh \hbar_{eff} {\D \E \over \E}
\eeq
the condition for coherent interference becomes
\beq
         {\d S \over \hbar_{eff} } < {\hbar_{eff} \over \d \hbar}
\label{dS}
\eeq

    The argument above is quite general, and applies
to any WKB treatment of the evolution equation \rf{Seq1}.
In fact, if one is prepared to accept that there may be
a stochastic uncertainty $\d \hbar$ (of whatever origin)
in the phenomenological value of Planck's
constant,  then a condition of the form \rf{dS} can be
easily deduced from the standard Feynman path integral in fixed background
spacetime.  If there are two or more semiclassical paths which contribute
to a given transition amplitude at leading order in $\hbar$, i.e.
\beq
      G[q_f,q_0] \approx \sum_{i} \m_i e^{iS_i[q_f,q_0]/\hbar}
\eeq
and if $\hbar$ itself has some dispersion $\d \hbar$, then the relative
phase between path $i$ and path $j$ becomes indeterminate
if the inequality \rf{dS} is violated, where
$\d S = |S_i-S_j|$ is the difference in action of the
two paths, and $\hbar_{eff}\equiv \hbar$.

   A signature of finite dispersion $\d \hbar$ in the effective
value of Planck's constant could be, e.g., an observed decoherence of
particle beams in an ultra-sensitive particle interferometer, in a
situation where standard time-energy considerations would imply that the
beams should interfere coherently.  In this case, the wavefunction $\Psi_A$
($\Psi_B$) represents the
contribution to the full ``wavefunction of the Universe'' $\Psi$ in which
the particle travels through path A (B) of the interferometer,
respectively, while $\d S / \hbar_{eff}$ is a WKB phase difference associated
with this difference in path. If the Universe
propagates slightly off-shell, as has been suggested here,  then
the interference will be incoherent if the inequality \rf{dS}
is violated.  To our knowledge no such decoherence has ever
been observed, and, in the absence of any theoretical lower bound
on $\d \hbar$, a more detailed discussion
of particle interferometry in this context would be premature.

   Of course, any finite dispersion in Planck's constant would also feed into
finite uncertainties in every other physical quantity, and some of these
quantities have been measured quite accurately.  In particular,
$\hbar / e^2$ can be
deduced, by combining $g-2$ measurements with high-order QED calculations,
to one part in $10^{12}$.  However, an ultra-high accuracy measurement
of some physical constant, such as $\hbar / e^2$, does not necessarily
project the Universe into an eigenstate of $\E$ (or $\hbar_{eff}$).
Planck's constant
is not determined from a single measurement (although $g-2$ {\it can}
be determined from observations of a single electron), and the
reported value would be, in our formalism, an average value
for $\hbar_{eff}$, at the average value $\E=\l^2$.  For example, in the
$g-2$ experiments, one adjusts a rf frequency to maximize the number
of spin flips of a trapped electron \cite{g-2}.  Naturally, the peak in
spin-flips versus frequency has a certain width.  The dispersion $\d \hbar$,
if indeed there is such a dispersion, would be a contribution
(perhaps negligible, compared to other sources) to that
width, while the center of the peak would locate, in the quantity
$\hbar / e^2$, only the average value of the effective Planck's constant.

\section{Conclusions}

   We have seen that the classical dynamics of bosonic fields (including
gravity) in a closed Universe can be re-expressed as describing the free fall
of a point particle in superspace.  The Hamiltonian operator describing this
``particle'' contains a (classically unobservable) parameter $\l$ analogous
to mass, and the usual Hamiltonian constraint of general relativity can
be viewed, in terms of this parameter, as a mass-shell condition.

   This ``free-fall'' description of general relativity is, of course,
a formal result.  Conceivably it also has physical content, and we have
suggested two possibilities:  First, quantum effects (virtual universe
loops) could induce an effective action for the (non-standard) supermetric,
and this action is essentially a function of the coupling constants
of the bosonic field theory.  In various minisuperspace models, we have seen 
that the effective action (or at least, the first terms in its adiabatic 
expansion) is stationary for vanishing cosmological constant.  We do not know
whether this desirable feature survives in the full theory.
Secondly, one may speculate that the universe, propagating like a
particle, may propagate slightly off-shell.  In principle this could lead
to some very interesting effects, as suggested in the last section, but
unfortunately we have no estimate to offer of their magnitude.

\vspace{33pt}

\noindent {\Large \bf Acknowledgements}

\bigskip

J.G.'s research is supported in part by  the U.S. Dept. of Energy, under
Grant No. DE-FG03-92ER40711.  A.C.'s research is supported by
a JSPS postdoctoral fellowship, under contract No. P95196; he would like
to thank the cosmology group at Tokyo Institute of Technology for the
kind hospitality during this work.  Support was also provided by the
Director, Office of Energy Research, Office of Basic Energy Services,
of the U.S. Department of Energy under Contract DE-AC03-76SF00098.

\newpage

\appendix
\section{$N_s=1$ massive scalar field}

In the case of a single massive, minimally coupled scalar field in a FRW
geometry, the effective action \rf{adiabat1} (neglecting contributions
coming from the supercurvature terms) reads
\bea
S_{eff}&\simeq &\L_S\int_0^{\ab} da\int_{-\phi_{M}}^{\phi_{M}} d\phi
~a^3 |(m^2\phi^2+\la)a^2-1|
\non \\
&=& {\L_{S}I_{\phi}\ab^4\over 36\xp}\biggl \{
[2(3\xp+\yp)-9]\xp
+{3\over (\xp+\yp)}
\non \\
&+&\Th(\xp){3\over (\xp\yp)^{1/2}}\arcsin\left[{\yp^{1/2}
\over(\xp+\yp)^{1/2}}\right ]
\non \\
&-&\Th(-\xp)\biggl [{(1-\xp)^{1/2}(8\xp^2-14\xp+3)\over \yp^{1/2}}
\non \\
&+&{3\over (-\xp\yp)^{1/2}}\ln\left[{[\yp^{1/2}+(-\xp)^{1/2}]^{1/2}\over
[\yp^{1/2}-(-\xp)^{1/2}]^{1/2}
[(-\xp)^{1/2}+(1-\xp)^{1/2}]}\right]\biggr]\biggr \}
\lb{r35}
\eea
where, in performing the integrations in $a$ and $\phi$, we have assumed that
the cutoff regulators satisfy the conditions $|\la|\ab^2 >1$
and $m^2\phi_M^2 >|\la|$, and we have defined, as usual, the variables
$\xp$ and $y$ according to eqs. \rf{g5} and \rf{f5}.
Taking the partial derivatives of $S_{eff}$ with respect to $\la$ and
$m^2$ we then get (for $x>0$)
\bea
{\partial S_{eff}\over \partial \la}&\simeq& {\L_{S}
I_{\phi}\ab^6\over 24}\biggl \{ {[4(\xp+\yp)^2\xp^2-(5\xp+3\yp)]\over
(\xp+\yp)^2}
\non \\
&-&{3\over (\xp\yp)^{1/2}}\arcsin\left[{\yp^{1/2}
\over(\xp+\yp)^{1/2}}\right ]\biggr \}
\non \\
{\partial S_{eff}\over \partial (m^2)}&\simeq& {\L_{S}
(I_{\phi})^3\ab^6\over 288\xp\yp}\biggl \{ {[4\xp\yp(\xp+\yp)^2+3(\xp-\yp)]
\over (\xp+\yp)^2}
\non \\
&-&{3\over (\xp\yp)^{1/2}}\arcsin\left[{\yp^{1/2}
\over(\xp+\yp)^{1/2}}\right ]\biggr \}
\lb{r40}
\eea
Summing and subtracting eqs. \rf{r40}, it is easy to check that the
stationarity conditions for the effective action become (for $x>0$)
\bea
2&\simeq&(\xp+\yp)^2(\xp-\yp)
\non \\
{\yp\over \xp+\yp}&\simeq&
\sin^2\left [{(\xp\yp)^{1/2}(3\xp^2+3\yp^2+2\xp\yp)\over 6}
\right ]
\lb{r42}
\eea
The method is to solve the first of conditions \rf{r42} for $\xp$ in terms
of $\yp$, and then to solve for $\yp$ from the second of \rf{r42}.
Although the functional dependence of $\xp$ on $\yp$ is unique, it
is easy to check that, already when $\xp>0$, there is an infinite set of
stationary points $y_{n}$,
with $x_{n}(y_{n})\sim \beta y_{n}\rightarrow +\infty$ as $n\rightarrow
\infty$.
In other words, the stationary points of $S_{eff}$ for $\xp>0$ will be at
\bea
\la_{n} &\sim &{\beta M_n\over \ab^2}
\non \\
m^2_n &\sim & {M_n \over \phi_{M}^2\ab^2}
\lb{r47}
\eea
Unfortunately, since $M_n\rightarrow\infty $ as $n\rightarrow \infty$,
one cannot make any reliable prediction (unless one assumes some - unnatural -
scaling between $M_n$ and the cutoff $\ab$) about the values of $\la$ and
$m$ in this toy model.

\end{document}